\documentclass[twocolumn,showpacs,amsmath,amssymb]{revtex4-1}
\usepackage{bm}
\usepackage[T1]{fontenc}
\usepackage{hyperref}
\usepackage{tabularx}
\input{epsf}

\usepackage{graphicx}
\usepackage{epstopdf}
\usepackage{array}
\usepackage{longtable}
\usepackage{rotating,booktabs}
\usepackage{booktabs,threeparttable}
\usepackage{bm}
\usepackage{float}
\usepackage{amsmath}
\usepackage{gensymb}
\usepackage{multirow}
\usepackage{epsfig}
\usepackage{color}

\begin{document}
\title{The critical role of negative-energy states in the Land\'{e} $g$-factor of lithium-like ions}
\author{Chang-Xian Song,$^{1}$ Yong-Bo Tang$^{1,*}$}

\affiliation {$^1$College of Engineering Physics, Shenzhen Technology University, Shenzhen 518118, China}

\email{tangyongbo@sztu.edu.cn}


\begin{abstract}
We report relativistic many-body calculations of the interelectronic-interaction correction to the Land\'{e} $g$-factor of the $2s_{1/2}$, $2p_{1/2}$, $2p_{3/2}$, and $3s_{1/2}$ states in lithium-like ions with nuclear charge $Z = 4-20$. Starting from the Dirac-Coulomb-Breit Hamiltonian, we treat positive-energy contributions using the coupled-cluster method with single and double excitations and include negative-energy contributions through third-order perturbation theory. We observe that negative-energy states give a state-dependent correction whose magnitude and sign vary with both Z and the state; for $2p_{1/2}$, the correction from the negative-energy states reaches 30\% of the total interelectronic-interaction contribution at $Z = 20$. Agreement with previous high-precision calculations is better than $0.1\%$, confirming the reliability of the present approach. This work may serve as a valuable reference for future precise calculations of $g$-factors for many-electron atomic systems.
\end{abstract}

\maketitle
\section{Introduction}
The Land\'{e} $g$-factor is a fundamental quantity that relates atomic magnetic moments to angular momentum. As a high-precision atomic observable, it serves as a sensitive probe for tests of bound-state quantum electrodynamics (QED) and for searches of physics beyond the Standard Model~\cite{Sturm.2025.V107.p023002,Akulov.2025.P52.v13,Heie.2023.V131.p25,Feynman.1949.V76.p769}. Moreover, it provides an important route for independent determinations of some fundamental constants, such as the electron mass~\cite{Sturm.2014.V506.p467,Zatorski.2017.V96.p012502} and the fine-structure constant~\cite{Shabaev.2006.V96.p253002}.
Recent advances in Penning-trap spectroscopy have pushed the measurement accuracy of $g$-factors for hydrogen-like ions to the order of $10^{-11}$~\cite{Morgner.2025.V134.p123201,Morgner.2023.P53.v622}, creating a stringent experimental benchmark for fundamental theory.

As a simple many-electron system, the $g$-factor of lithium-like ions has attracted considerable interest over the past two decades~\cite{Yan.2002.V35.p1885,Yan.2002.V66.p,Glazov.2004.V70.p062104,Florian.2016.V7.p10246,Yerokhin.2016.V94.p022502,Yerokhin.2017.V95.p062511,Glazov.2019.V123.p173001,Yerokhin.2020.V102.p022815,Yerokhin.2021.V104.p022814,Kosheleva.2022.V128.p103001,zinenko.2025.P.v}. For the ground state, experimental precision has reached the $10^{-10}$ level, comparable to that of hydrogen-like systems, while theoretical accuracy remains somewhat lower~\cite{Kosheleva.2022.V128.p103001}. For excited states, experimental data are still absent, and theoretical predictions lag far behind those of hydrogen-like systems~\cite{zinenko.2025.P.v,Morgner.2025.V134.p123201}. The primary obstacle is the complexity of many-electron correlation effects, which demand a simultaneous, high-accuracy treatment of interelectronic-interaction corrections, higher-order relativistic corrections, and many-body QED effects.

Since the systematic framework was established by Glazov $et~al$.~\cite{Glazov.2004.V70.p062104}, the theoretical description of the ground-state $g$-factor of lithium-like ions has matured considerably~\cite{Moskovkin.2008.V77.p063421,Wagner.2013.V110.p033003,Volotka.2014.V112.p253004,Glazov.2019.V123.p173001,Kosheleva.2022.V128.p103001,Yerokhin.2020.V102.p022815,Yerokhin.2021.V104.p022814,Kosheleva.2022.V128.p103001}. Nonperturbative QED calculations have achieved precision comparable to experiment over a wide range of nuclear charge numbers~\cite{Kosheleva.2022.V128.p103001,Yerokhin.2021.V104.p022814,Glazov.2019.V123.p173001}. By contrast, studies of the $2p$ excited states remain sparse and significantly less accurate~\cite{Guan.1998.V244.p120,Yan.2002.V66.p,zinenko.2025.P.v}. In recent years, experimental efforts have shifted toward the $g$-factors of "non-$S$" states, as demonstrated by high-precision measurements in boron-like Ar ions (the $1s^2 2s^2 2p_{1/2,3/2}$ states)~\cite{Arapoglou2019PRL,Micke2020Nature}, carbon-like Ca ions (the $1s^2 2s^2 2p^2\;{}^3P_1$ state)~\cite{Ca_PRL2025}, and $^{48}$Ti$^+$ ions (the $3d^2 4s\;{}a^4F_{3/2-9/2}$ and $3d^3\;{}b^4F_{9/2}$ states)~\cite{TiPRL}. Although these results have not yet reached the benchmark precision achieved for hydrogen-like and lithium-like ground states~\cite{Morgner.2025.V134.p123201,Morgner.2023.P53.v622,Glazov.2004.V70.p062104,zinenko.2025.P.v}, they are sufficiently accurate to test relativistic many-body atomic theory and some physical effects. These "non-$S$" states are particularly sensitive to many-body electron correlations; achieving high-precision $g$-factors relies critically on accurate quantification of interelectronic-interaction corrections. A key aspect that has not been adequately addressed in existing theoretical treatments is the contribution of negative-energy states~\cite{Yan.2002.V66.p}. Physically, negative-energy states correspond to the creation of virtual electron-positron pairs in the Dirac sea~\cite{Jiang.2025.V112.p062223}. Their contributions are partially screened in $s$ states~\cite{Surzhykov.2009.V80.p052511}. However, in $2p$ states, owing to the lower wavefunction density near the nucleus, relativistic-correlation cross terms may amplify the influence of the negative-energy continuum~\cite{Surzhykov.2009.V80.p052511}, calling for dedicated quantitative analysis.

Early calculations for the $2p$ states of lithium-like ions, such as those by Yan $et~al.$~\cite{Yan.2002.V66.p} using Hylleraas-type wavefunctions, entirely neglected negative-energy contributions. Zinenko $et~al.$~\cite{zinenko.2025.P.v} proposed the first complete relativistic treatment of the lithium-like
$g$-factor and incorporated the negative-energy states within the QED framework. However, the interelectronic-interaction correction was treated only to second order, and contributions from third and higher orders rely on estimates based on a parameterized screening potential, whose accuracy warrants further independent assessment. Moreover, the framework did not separate or quantitatively analyze the specific contribution from the negative-energy continuum.
The recent study by Wang $et~al.$~\cite{Wang.2025.P109413.v388} on boron-like ions employed many-body perturbation theory to explicitly demonstrate that the negative-energy contribution can be as significant as other correlation corrections.

In this work, we perform high-precision calculations of the interelectronic-interaction correction to the $g$-factors for lithium-like ions with nuclear charge numbers $Z = 4$-$20$, considering the ground state $2s_{1/2}$ and the excited states $2p_{1/2}$, $2p_{3/2}$, and $3s_{1/2}$. The dependence of the negative-energy contribution on $Z$ is systematically investigated. Our approach combines relativistic coupled-cluster theory in the positive-energy correlations and third-order many-body perturbation theory for the negative-energy contribution.
In contrast to the screening-potential approximation of Zinenko $et~al.$~\cite{zinenko.2025.P.v}, our coupled-cluster framework treats electron correlation nonperturbatively while extracting the negative-energy contribution quantitatively via third-order perturbation.
Compared to the many-body perturbation theory employed by Wang $et~al.$~\cite{Wang.2025.P109413.v388} for boron-like ions, our method achieves a higher degree of nonperturbative completeness in the treatment of positive-energy correlations. Through this unified scheme, we obtain $g$-factor predictions for many-electron systems with an accuracy comparable to that of dedicated few-body methods, and we present a systematic, quantitative study of the state-dependent contributions arising from negative-energy states.

The paper is organized as follows. Section II details the theoretical methodology. The computational details and results are presented and discussed in Section III, which includes a comprehensive analysis of the interelectronic-interaction correction and the contributions from negative-energy states. Finally, a summary and outlook are provided in Section IV.

\vspace{-0.5cm}

\section{Theoretical method}\label{sec:theoretical}
\subsection{Land\'{e} $g$-factor}
The interaction between the atomic magnetic moment and an external magnetic field gives rise to the Zeeman effect. The Land\'{e} $g$-factor quantifies the magnitude of the resulting level splitting and is defined as the ratio of the projection of the total atomic magnetic moment along the direction of the total angular momentum to the Bohr magneton $\mu_B$. It can be expressed as
\begin{equation}
 g_{J}=\frac{1}{2 \mu_{B}} \frac{< \Psi_{v}\| \bm N^{(1)}\| \Psi_{v}>}{\sqrt{J(J+1)(2 J+1)}},
\end{equation}
where the tensor operator $N^{(1)}$ is given by
\begin{equation}
\bm N^{(1)}=-i \sum_{j} \sum_{q=0, \pm 1} \sqrt{\frac{8 \pi}{3}} r_{j} \bm \alpha_{j} \cdot \bm Y_{1 q}^{(0)}\left(\hat{\bm r}_{j}\right)\ .
\end{equation}
Here, $\bm \alpha = \bigl(\begin{smallmatrix} 0 & \sigma \\ \sigma & 0 \end{smallmatrix}\bigr)$ represents the vector spherical harmonic~\cite{Cheng.1985.V31.p2775}.

The QED correction to the $g$-factor takes the form
\begin{equation}
\Delta g=\frac{\left(g_{s}-2\right)}{2}\frac{\left<\Psi_{v}\left\| \Delta \bm{N} ^{(1)}\right\| \Psi_{v}\right>}{ \sqrt{J(J+1)(2 J+1)}}, \quad
\end{equation}
where the electron spin $g$-factor $g_s$ is
\begin{equation}
g_{s}=2\left[1+\frac{\alpha}{2 \pi}-0.328 \frac{\alpha^{2}}{\pi^{2}}+\cdots\right] \approx 2.0023193 ,
\end{equation}
and the corresponding operator is defined as
\begin{equation}
\Delta \bm N^{(1)}=\bm \sum_{j}^{N} \bm \beta_{j} \bm \sum_{j}.
\end{equation}
In this expression, $\bm \beta$  is the standard Dirac matrix, and $\bm \sum$ is the relativistic spin matrix.
The total Land\'{e} $g$-factor is therefore given by the sum
\begin{equation}
g = g_J + \Delta g.
\end{equation}

For few-electron ions, the theoretical $g$-factor is evaluated within the Dirac framework by systematically including higher-order corrections~\cite{Shabaev.2002.V65.p062104}
\begin{equation}
g = g_{\text{D}} + \Delta g_{\text{int}} + \Delta g_{\text{QED}} + \Delta g_{\text{nuc}}.
\end{equation}
Here, $g_{\text{D}}$ is the Dirac value, representing the $g$-factor of an electron moving in the Coulomb potential of the nucleus; $\Delta g_{\text{int}}$ denotes interelectronic-interaction correction; $\Delta g_{\text{QED}}$ represents QED correction; and $\Delta g_{\text{nuc}}$ accounts for nuclear effects, such as finite nuclear size and nuclear polarization. The primary focus of this work is calculation of $\Delta g_{\text{int}}$. The QED correction $\Delta g_{\text{QED}}$ and the nuclear correction $\Delta g_{\text{nuc}}$ are calculated within the framework of rigorous quantum electrodynamics theory, following established methods cited in the literature~\cite{Glazov.2004.V70.p062104,Moskovkin.2008.V77.p063421,Kosheleva.2022.V128.p103001,zinenko.2025.P.v}.

For lithium-like ions with zero nuclear spin, the Dirac $g$-factors for the ground state and low-lying excited states are given by the following analytical expressions
\begin{align}
g_{D}(2s_{1/2},\,3s_{1/2}) &= \frac{2}{3}\left[1+\sqrt{2+2\sqrt{1-(\alpha Z)^{2}}}\right],\\
g_{D}(2p_{1/2}) &= \frac{2}{3}\left[\sqrt{2\left(1+\sqrt{1-(\alpha Z)^{2}}\right)}-1\right],\\
g_{D}(2p_{3/2}) &= \frac{4}{15}\left[2\sqrt{4-(\alpha Z)^{2}}+1\right],
\end{align}
where $\alpha = 1/137.035\,999\,074$ is the fine-structure constant.

\subsection{ Relativistic many-body approach}
In lithium-like ions, which are monovalent atomic systems, the exact atomic wave function $|\Psi_{v}\rangle$ for a valence orbital $v$ is expressed using a wave operator $\Omega$ acting on a reference state
\begin{equation}\label{EQ:ASF}
|\Psi_{v}\rangle = \Omega\,|\Phi_{v}\rangle,
\end{equation}
where $|\Phi_{v}\rangle$ is the Dirac--Fock (DF) wave function, serving as the zeroth-order reference, and $\Omega$ is the wave operator. There are two distinctive methods to express the wave operator: many-body perturbation theory (MBPT)~\cite{Tang.2019.V52.p055002} and the coupled-cluster method~\cite{Lindgren.1978.V14.p33,Tang.2017.V96.p022513}.

Within the framework of many-body perturbation theory (MBPT), the wave operator is expanded as
\begin{equation}
\Omega = \sum_{n=0}^{\infty} \Omega^{(n)},
\end{equation}
where $n$ denotes the perturbation order. The $n$th-order correction to the wave function is given by
\begin{equation}
|\Psi_{v}^{(n)}\rangle = \Omega^{(n)}\,|\Phi_{v}\rangle .
\end{equation}
The expectation value of an operator $\hat{O}$ for a state $v$, correct to $n$th order, is given by
\begin{equation}
\overline{O}_{v}^{(n)} =
\frac{\langle\Phi_{v}|\sum_{i,j=0}^{i+j<n} \Omega^{(i)\dagger}\hat{O}\,\Omega^{(j)}|\Phi_{v}\rangle}
     {\langle\Phi_{v}|\sum_{i,j=0}^{i+j<n} \Omega^{(i)\dagger}\Omega^{(j)}|\Phi_{v}\rangle}.
\end{equation}
The computational complexity grows rapidly with the order $n$.

In this work, the perturbation expansion is truncated at third order, yielding
\begin{equation}
\overline{O}_{v} \approx \overline{O}_{v}^{(1)} + \overline{O}_{v}^{(2)} + \overline{O}_{v}^{(3)}.
\end{equation}
The first-order term is simply the DF expectation value
\begin{equation}
\overline{O}_{v}^{(1)} = \langle\Phi_{v}|\hat{O}|\Phi_{v}\rangle.
\end{equation}
The second-order term is expressed as
\begin{equation}
\begin{aligned}
\overline{O}_{v}^{(2)}
&= \langle\Phi_{v}|\Omega^{(1)\dagger}\hat{O} + \hat{O}\Omega^{(1)}|\Phi_{v}\rangle \\
&= \sum_{a r}\frac{O_{ar}(g_{v r v a}-g_{v r a v})}{\varepsilon_{a}-\varepsilon_{r}}
 + \sum_{a r}\frac{(g_{v a v r}-g_{v a r v})O_{ra}}{\varepsilon_{a}-\varepsilon_{r}} \\
&= \sum_{a r}\frac{O_{ar}\tilde{g}_{v r v a}}{\varepsilon_{a}-\varepsilon_{r}}
 + \sum_{a r}\frac{\tilde{g}_{v a v r}O_{ra}}{\varepsilon_{a}-\varepsilon_{r}},
\end{aligned}
\end{equation}
where $a$ denotes core orbitals and $r$ virtual orbitals, which can simultaneously include the negative-energy states ($\mathcal{N}$) and the positive-energy states ($\mathcal{P}$). When the Coulomb and Breit corrections are included, the electron--electron interaction matrix element $g_{ijkl}$ is replaced by $\tilde{g}_{ijkl}=g_{ijkl}-g_{ijlk}$. $\varepsilon_{i}$ is the DF energy of orbital $i$.

The third-order expectation value is given by
\begin{equation}
\overline{O}_{v}^{(3)} =
\frac{\langle\Phi_{v}|\Omega^{(2)\dagger}\hat{O} + \hat{O}\Omega^{(2)} + \Omega^{(1)\dagger}\hat{O}\Omega^{(1)}|\Phi_{v}\rangle}
     {\langle\Phi_{v}|1 + \Omega^{(1)\dagger}\Omega^{(1)}|\Phi_{v}\rangle}.
\end{equation}

\begin{table*}
\caption{Convergence of the interelectronic-interaction corrections from negative-energy states $\Delta g_{\rm{int}}$ ($\mathcal{N}$) and positive-energy states ($\Delta g_{\mathrm{int}}^{\rm{MBPT}}(\mathcal{P})$, $\Delta g_{\mathrm{int}}^{\rm{LCCSD}}(\mathcal{P})$, $\Delta g_{\mathrm{int}}^{\rm{CCSD}}(\mathcal{P})$) with respect to partial wave $\ell$ in different states of lithium-like Ca$^{17+}$ ions, where $\Delta g_{\mathrm{int}}^{\rm{MBPT}}(\mathcal{P})$, $\Delta g_{\mathrm{int}}^{\rm{LCCSD}}(\mathcal{P})$ and $\Delta g_{\mathrm{int}}^{\rm{CCSD}}(\mathcal{P})$ are calculated via the MBPT, LCCSD, and CCSD methods, respectively. The values in parentheses indicate the uncertainties.
 \label{conver}}
\begin{ruledtabular}
\scriptsize
\begin{tabular}{llllllllllllrrrrrll}
$\ell_{\rm{max}}$  &  $\Delta g_{\mathrm{int}}(\mathcal{N})$ &$\Delta g_{\mathrm{int}}^{\rm{MBPT}}(\mathcal{P})$ & $\Delta g_{\mathrm{int}}^{\rm{LCCSD}}(\mathcal{P})$ & $\Delta g_{\mathrm{int}}^{\rm{CCSD}}(\mathcal{P})$ & $\Delta g_{\mathrm{int}}(\mathcal{N})$  & $\Delta g_{\mathrm{int}}^{\rm{MBPT}}(\mathcal{P})$ & $\Delta g_{\mathrm{int}}^{\rm{LCCSD}}(\mathcal{P})$ & $\Delta g_{\mathrm{int}}^{\rm{CCSD}}(\mathcal{P})$ \\
\hline
 &  &  & $2s_{1/2}$                                               & & & &$3s_{1/2}$\\
2	          & $-$0.0000277801 &   0.0004822099& 0.0004819860     &0.0004820026    &$-$0.0000078992&	0.0022337979 	  &0.0022337384  	&0.0022337427    \\
3	          & $-$0.0000277853 &   0.0004821498& 0.0004819289     &0.0004819459    &$-$0.0000079014&	0.0022337832 	  &0.0022337246  	&0.0022337290    \\
4	          & $-$0.0000277866 &   0.0004821275& 0.0004819081     &0.0004819254    &$-$0.0000079018&	0.0022337777 	  &0.0022337196  	&0.0022337240    \\
5	          & $-$0.0000277868 &   0.0004821174& 0.0004818989     &0.0004819162    &$-$0.0000079018&	0.0022337752 	  &0.0022337173  	&0.0022337217    \\
6	          & $-$0.0000277868 &   0.0004821122& 0.0004818942     &0.0004819116    &$-$0.0000079018&	0.0022337739 	  &0.0022337161  	&0.0022337206    \\
Extrapolated&						      &               &                  &                &               &                 &               &              \\
2-3-4-5	    &$-$0.0000277855 	&   0.0004821032& 0.0004818862     &0.0004819036    &$-$0.0000079018  &	0.0022337716 	&0.0022337141  	&0.0022337186    \\
3-4-5-6	    &$-$0.0000277863 	&   0.0004821032& 0.0004818863     &0.0004819037    &$-$0.0000079016  &	0.0022337716 	&0.0022337141  	&0.0022337186    \\
Finally	    &$-$0.00002779(1)	&  0.00048211(1)&	0.00048189(1)	   &0.00048191(1)	  &$-$0.00000790(1) &	0.00223377(1)	&0.00223371(1)	&0.00223372(1)  \\
	          &                 &               &$2p_{1/2}$		    &                  &                &  & $2p_{3/2}$		      &                 &                     \\
2	          & $-$0.000170797	&   0.000767515 &	 0.000767471 	   &  0.000767480 	& 0.000014941 	&  0.000469942 	  &0.000469933   	&0.000469928    \\
3	          & $-$0.000170735	&   0.000767551 &	 0.000767508 	   &  0.000767518 	& 0.000015007 	&  0.000469750 	  &0.000469746   	&0.000469741    \\
4	          & $-$0.000170723	&   0.000767557 &	 0.000767515 	   &  0.000767524 	& 0.000015018 	&  0.000469694 	  &0.000469693   	&0.000469688    \\
5	          & $-$0.000170718	&   0.000767560 &	 0.000767517 	   &  0.000767527 	& 0.000015024 	&  0.000469671 	  &0.000469672   	&0.000469667    \\
6	          & $-$0.000170715	&   0.000767561 &	 0.000767518 	   &  0.000767528 	& 0.000015027 	&  0.000469661 	  &0.000469662   	&0.000469657    \\
Extrapolated&						      &               &                  &                &               &                 &               &             \\
2-3-4-5	    &$-$0.000170700 	&   0.000767566 &	 0.0007675239    &  0.0007675327 	& 0.0000150483 	&  0.0004696399 	&0.0004696420 	&0.0004696369   \\
3-4-5-6	    &$-$0.000170711	  & 0.000767564 	&  0.0007675199 	 &  0.0007675298 	& 0.0000150308 	&  0.0004696446 	&0.0004696460 	&0.0004696410   \\
Finally	    &$-$0.00017071(1) &	0.00076756(1)	&  0.00076752(1)   &	0.00076753(1)	&0.00001503(1)	&  0.00046965(1)	&0.00046965(1)	&0.00046965(1)  \\
\end{tabular}
\end{ruledtabular}
\end{table*}

In the coupled-cluster (CC) formalism~\cite{Lindgren.1978.V14.p33,Tang.2017.V96.p022513}, the wave operator $\Omega$ is expressed as an exponential of the cluster operator $S$
\begin{equation}
    \Omega = e^{S}.
\end{equation}
The cluster operator $S$ is partitioned into many-body components
\begin{equation}
    S=\sum_{m=1}^{M}S_{m} = S_1+S_2+S_3+\dots+S_m,
\end{equation}
where $M$ is the total number of electrons and $S_m$ is the $m$-electron excitation operator. When the expansion is truncated at single and double excitations ($S \approx S_1 + S_2$), the method is termed as CCSD. The corresponding wave function is given by $\left|\Psi_{v}\right>_{\mathrm{CCSD}}= e^{S_1+S_2}\left|\Phi_{v}\right>$. Expanding the exponential yields a series of terms
\begin{equation}\label{EQ:CCSD}
 \begin{aligned}
 \left|\Psi_{v}\right>_{\mathrm{CCSD}}
 ={}& \bigl\{1+S_{1}+S_{2}+\tfrac{1}{2!}\bigl(S_{1}^{2}+S_{2}^{2}+S_{1} S_{2}\bigr) \\
 &+\tfrac{1}{3!}\bigl(S_{1}^{3}+3 S_{1}^{2} S_{2}\bigr)+\tfrac{1}{4!} S_{1}^{4}\bigr\}\left|\Phi_{v}\right>.
 \end{aligned}
\end{equation}
In the CCSD wave-function expansion, the linear terms $S_1$ and $S_2$ correspond to all single- and double-excitation configurations, respectively, and constitute the dominant part of the electron-correlation correction. The nonlinear terms (such as $S_1^2$, $S_1S_2$, etc.) describe the coupling between excitation operators and are essential for preserving the size extensivity of the theory. However, the computational cost of these higher-order terms is significantly greater than that of the linear terms.

When correlation effects are weak or computational resources are limited, a practical approximation is to retain only the most important linear terms while neglecting the higher-order coupling terms. This defines the linear coupled-cluster method (LCCSD)
\begin{equation}\label{EQ:LCCSD}
 \left|\Psi_{v}\right>_{\mathrm{LCCSD}}\approx  \left\{1+S_{1}+S_{2}\right\}\left|\Phi_{v}\right> .
\end{equation}
While sacrificing some theoretical rigor, LCCSD substantially reduces computational cost and is often employed as an effective tool for studying weakly correlated systems, or as an initial step in iteratively solving the full CCSD equations.

For property calculations within this framework, such as the evaluation of the $g$-factor, the coupled-cluster approach differs fundamentally from the MBPT introduced earlier.
The expectation value of an operator $\hat{O}$ for state $\left|\Psi_v\right>$ in coupled-cluster theory is evaluated as
\begin{equation}\label{eq:cc_exp_val}
\overline{O_v}=
\frac{\left<\Psi_{v}\right|\hat{O}\left|\Psi_{v}\right>}
     {\left<\Psi_{v}|\Psi_{v}\right>}
=\frac{\left<\Phi_{v}\right|e^{S^{\dagger}}\hat{O}e^{S}\left|\Phi_{v}\right>}
      {\left<\Phi_{v}\right|e^{S^{\dagger}}e^{S}\left|\Phi_{v}\right>},
\end{equation}
where $S^{\dagger}$ denotes the Hermitian conjugate of $S$. The expansion of this expression leads to an infinite series. In practice, by applying the LCCSD approximation from Eq.~\eqref{EQ:LCCSD}, the numerator and denominator in Eq.~\eqref{eq:cc_exp_val} can be expanded as follows
\begin{equation}
\begin{aligned}
e^{S^{\dagger}}\hat{O}e^{S}
\approx{}& \hat{O}
+\bigl\{\hat{O} S_{1}^{(0,0)\dagger}+ \text{c.c.}\bigr\}
+\bigl\{\hat{O} S_{1}^{(0,1)}+ \text{c.c.}\bigr\} \\
&+\bigl\{\hat{O} S_{2}^{(0,1)}+ \text{c.c.}\bigr\}
+\bigl\{S_{1}^{(0,0)\dagger}\hat{O} S_{1}^{(0,1)}+ \text{c.c.}\bigr\} \\
&+S_{1}^{(0,0)\dagger}\hat{O} S_{1}^{(0,0)}
+\bigl\{S_{1}^{(0,0)\dagger}\hat{O} S_{2}^{(0,0)}+ \text{c.c.}\bigr\} \\
&+\bigl\{S_{1}^{(0,0)\dagger}\hat{O} S_{2}^{(0,1)}+ \text{c.c.}\bigr\}
+S_{2}^{(0,0)\dagger}\hat{O} S_{2}^{(0,0)} \\
&+\bigl\{S_{2}^{(0,0)\dagger}\hat{O} S_{2}^{(0,1)}+ \text{c.c.}\bigr\}
+S_{1}^{(0,1)\dagger}\hat{O} S_{1}^{(0,1)} \\
&+\bigl\{S_{1}^{(0,1)\dagger}\hat{O} S_{2}^{(0,1)}+ \text{c.c.}\bigr\}
+S_{2}^{(0,1)\dagger}\hat{O} S_{2}^{(0,1)}
\end{aligned}
\end{equation}
and
\begin{equation}
\begin{aligned}
e^{S^{\dagger}}e^{S}
\approx{}& 1
+S_{1}^{(0,0)\dagger}S_{1}^{(0,0)}
+S_{1}^{(0,1)\dagger}S_{1}^{(0,1)} \\
&+S_{2}^{(0,0)\dagger}S_{2}^{(0,0)}
+S_{2}^{(0,1)\dagger}S_{2}^{(0,1)}.
\end{aligned}
\end{equation}
Here, ``$\text{c.c.}$'' denotes the complex conjugate. The superscripts in $S_{1}^{(0,0)}$ and $S_{1}^{(0,1)}$ denote excitations from core and valence electrons, respectively. The corresponding expansion under the full CCSD approximation involves many additional terms. In the present CCSD and LCCSD calculations, only positive-energy states are retained.

\subsection{Computation details}
The single-particle basis set is obtained by self-consistent solution of the Dirac--Fock equations based on the Dirac--Coulomb--Breit Hamiltonian. The radial wavefunctions are expanded using 40 B-spline functions of order $k = 11$, and the partial-wave expansion is truncated at $\ell_{\text{max}} = 6$. For each relativistic angular symmetry $\kappa$, this yields a discrete spectrum containing 40 positive-energy and 40 negative-energy eigenstates.

\begin{table*}
\caption{The interelectronic-interaction corrections from the positive-energy state $\Delta g_{\mathrm{int}}$ ($\mathcal{P}$) for the ground state and low-lying excited states of lithium-like ions ($Z = 4 - 20$), calculated using the MBPT, LCCSD, and CCSD methods, along with the final recommended values. Values in parentheses indicate the uncertainties.
 \label{int_Positive}}
\begin{ruledtabular}
\begin{tabular}{llllllllllllllll}
Z  &  $\Delta g_{\mathrm{int}}^{\rm{MBPT}}(\mathcal{P})$ & $\Delta g_{\mathrm{int}}^{\rm{LCCSD}}(\mathcal{P})$ & $\Delta g_{\mathrm{int}}^{\rm{CCSD}}(\mathcal{P})$ &  Recom. & $\Delta g_{\mathrm{int}}^{\rm{MBPT}}(\mathcal{P})$ & $\Delta g_{\mathrm{int}}^{\rm{LCCSD}}(\mathcal{P})$ & $\Delta g_{\mathrm{int}}^{\rm{CCSD}}(\mathcal{P})$ &  Recom.\\
\hline
 &  & & $2s_{1/2}$                                               & & &$3s_{1/2}$\\
4  & 0.00008800  &	0.00008744 &	0.00008750 &	0.00008750(50) & 0.00012143 	& 0.00012131 &	0.00012132 &	0.00012132(11) \\
5  & 0.00011258  &	0.00011203 &	0.00011208 &	0.00011208(50) & 0.00017858 	& 0.00017845 &	0.00017846 &	0.00017846(12) \\
6  & 0.00013710  &	0.00013658 &	0.00013663 &	0.00013663(47) & 0.00024558 	& 0.00024546 &	0.00024547 &	0.00024547(12) \\
7  & 0.00016162  &	0.00016114 &	0.00016114 &	0.00016114(48) & 0.00032241 	& 0.00032229 &	0.00032230 &	0.00032230(11) \\
8  & 0.00018614  &	0.00018570 &	0.00018574 &	0.00018574(41) & 0.00040911 	& 0.00040900 &	0.00040901 &	0.00040901(10) \\
9  & 0.00021069  &	0.00021028 &	0.00021032 &	0.00021032(38) & 0.00050601 	& 0.00050590 &	0.00050591 &	0.00050591(10) \\
10 & 0.00023526  &	0.00023488 &	0.00023491 &	0.00023491(35) & 0.00061264 	& 0.00061254 &	0.00061255 &	0.00061255(9 ) \\
11 & 0.00025984  &	0.00025949 &	0.00025951 &	0.00025951(33) & 0.00072924 	& 0.00072915 &	0.00072916 &	0.00072916(8 ) \\
12 & 0.00028444  &	0.00028411 &	0.00028414 &	0.00028414(31) & 0.00085591 	& 0.00085583 &	0.00085584 &	0.00085584(8 ) \\
13 & 0.00030907  &	0.00030876 &	0.00030878 &	0.00030878(29) & 0.00099282 	& 0.00099273 &	0.00099274 &	0.00099274(7 ) \\
14 & 0.00033372  &	0.00033342 &	0.00033344 &	0.00033344(27) & 0.00113953 	& 0.00113945 &	0.00113945 &	0.00113945(7 ) \\
15 & 0.00035838  &	0.00035811 &	0.00035813 &	0.00035813(26) & 0.00129634 	& 0.00129627 &	0.00129628 &	0.00129628(7 ) \\
16 & 0.00038308  &	0.00038281 &	0.00038283 &	0.00038283(24) & 0.00146343 	& 0.00146336 &	0.00146336 &	0.00146336(6 ) \\
17 & 0.00040780  &	0.00040755 &	0.00040756 &	0.00040756(23) & 0.00164059 	& 0.00164053 &	0.00164053 &	0.00164053(6 ) \\
18 & 0.00043254  &	0.00043232 &	0.00043232 &	0.00043232(22) & 0.00182813 	& 0.00182807 &	0.00182807 &	0.00182807(6 ) \\
19 & 0.00045731  &	0.00045709 &	0.00045710 &	0.00045710(21) & 0.00202590 	& 0.00202584 &	0.00202584 &	0.00202584(6 ) \\
20 & 0.00048211  &	0.00048189 &	0.00048191 &	0.00048191(20) & 0.00223377 	& 0.00223372 &	0.00223372 &	0.00223372(5 ) \\
\hline
 & & &$2p_{1/2}$                                                & &  &$2p_{3/2}$\\
4  & 0.00011198  &	0.00011184 &	0.00011185 & 0.00011185(13) & 0.00008097 	& 0.00008097 &	0.00008096   &  0.00008096(7)  \\
5  & 0.00015249  &	0.00015233 &	0.00015235 & 0.00015235(15) & 0.00010550 	& 0.00010551 &	0.00010550   &  0.00010550(8)  \\
6  & 0.00019318  &	0.00019302 &	0.00019304 & 0.00019304(14) & 0.00012982 	& 0.00012983 &	0.00012982   &  0.00012982(8)  \\
7  & 0.00023393  &	0.00023377 &	0.00023379 & 0.00023379(13) & 0.00015405 	& 0.00015406 &	0.00015406   &  0.00015406(8)  \\
8  & 0.00027470  &	0.00027456 &	0.00027458 & 0.00027458(12) & 0.00017825 	& 0.00017826 &	0.00017826   &  0.00017826(8)  \\
9  & 0.00031550  &	0.00031537 &	0.00031539 & 0.00031539(11) & 0.00020244 	& 0.00020245 &	0.00020244   &  0.00020244(8)  \\
10 & 0.00035633  &	0.00035621 &	0.00035622 & 0.00035622(10) & 0.00022662 	& 0.00022664 &	0.00022663   &  0.00022663(7)  \\
11 & 0.00039720  &	0.00039709 &	0.00039710 & 0.00039710(10) & 0.00025082 	& 0.00025083 &	0.00025082   &  0.00025082(7)  \\
12 & 0.00043810  &	0.00043800 &	0.00043802 & 0.00043802(9)  & 0.00027503 	& 0.00027504 &	0.00027503   &  0.00027503(7)  \\
13 & 0.00047906  &	0.00047897 &	0.00047898 & 0.00047898(8)  & 0.00029926 	& 0.00029927 &	0.00029926   &  0.00029926(7)  \\
14 & 0.00052007  &	0.00051998 &	0.00052000 & 0.00052000(7)  & 0.00032351 	& 0.00032352 &	0.00032351   &  0.00032351(6)  \\
15 & 0.00056114  &	0.00056106 &	0.00056107 & 0.00056107(7)  & 0.00034779 	& 0.00034779 &	0.00034779   &  0.00034779(6)  \\
16 & 0.00060227  &	0.00060220 &	0.00060221 & 0.00060221(6)  & 0.00037209 	& 0.00037210 &	0.00037209   &  0.00037209(6)  \\
17 & 0.00064347  &	0.00064341 &	0.00064342 & 0.00064342(5)  & 0.00039643 	& 0.00039644 &	0.00039643   &  0.00039643(6)  \\
18 & 0.00068475  &	0.00068471 &	0.00068471 & 0.00068471(5)  & 0.00042080 	& 0.00042080 &	0.00042080   &  0.00042080(2)  \\
19 & 0.00072611  &	0.00072606 &	0.00072607 & 0.00072607(4)  & 0.00044521 	& 0.00044521 &	0.00044521   &  0.00044521(5)  \\
20 & 0.00076756  &	0.00076752 &	0.00076753 & 0.00076753(3)  & 0.00046966 	& 0.00046966 &	0.00046966   &  0.00046966(5)  \\
\end{tabular}
\end{ruledtabular}
\end{table*}

Contributions of electron correlation to the Land\'e $g$-factor are evaluated using three independent many-body methods: third-order many-body perturbation theory (MBPT), the linearized coupled-cluster method with singles and doubles (LCCSD), and the full coupled-cluster method with singles and doubles (CCSD). To separately determine the contribution arising from negative-energy states ($\mathcal{N}$), a specific comparison is carried out within the MBPT framework. Two complete MBPT calculations are performed: one allowing virtual excitations to both positive- and negative-energy orbitals, giving $g_{\rm J}(\mathcal{P,N})$, and another restricted to only positive-energy orbitals, giving $g_{\rm J}(\mathcal{P})$. Their difference defines the negative-energy state correction,
\begin{equation}
\Delta g_{\rm int}(\mathcal{N}) = g_{\rm J}(\mathcal{P,N}) - g_{\rm J}(\mathcal{P}).
\end{equation}
The positive-energy correction is defined relative to the Dirac value $g_{\rm D}$,
\begin{equation}
\Delta g_{\rm int}(\mathcal{P}) = g_{\rm J}(\mathcal{P}) - g_{\rm D},
\end{equation}
with both Coulomb and Breit interactions included consistently. The LCCSD and CCSD calculations are conducted in the positive-energy virtual space. The total interelectronic-interaction corrections from these methods are then formed by adding the MBPT derived negative-energy contribution,
\begin{equation}
\Delta g_{\rm int}^{\rm LCCSD/CCSD} = \Delta g_{\rm int}^{\rm LCCSD/CCSD}(\mathcal{P}) + \Delta g_{\rm int}(\mathcal{N}).
\end{equation}

The convergence of the many-body results with respect to the partial-wave expansion is examined. To approach the complete basis-set limit ($\ell_{\text{max}} \rightarrow \infty$), an extrapolation is applied. The incremental correction $\Delta g_{\ell-(\ell-1)}$ follows an asymptotic form,
\begin{equation}\label{extrapolate}
\Delta g_{\ell-(\ell-1)} = \frac{C_{1}}{\ell^{4}} + \frac{C_{2}}{\ell^{5}} + \frac{C_{3}}{\ell^{6}},
\end{equation}
where the parameters $C_1$, $C_2$, and $C_3$ are fitted using results from four successive $\ell_{\text{max}}$ values, following the procedure in Ref.~\cite{Wang.2025.P109413.v388}. Direct calculations are performed up to $\ell_{\text{max}} = 6$. The stability of the extrapolation is checked by comparing results from two independent sequences of $\ell_{\text{max}}$ values, confirming the consistency of the final values.

This work concentrates on the interelectronic-interaction contributions. For a full theoretical prediction, higher-order quantum electrodynamic (QED) corrections and nuclear effects (size and polarization) are also required. These terms are taken from high-precision, few-body QED calculations reported in the literature~\cite{Glazov.2004.V70.p062104,zinenko.2025.P.v}. In the presentation of the final results, our calculated values are combined with these external QED and nuclear corrections.

\begin{table*}
\caption{Interelectronic-interaction corrections $\Delta g_{\mathrm{int}}$ for lithium-like ions ($Z = 4$-$20$), including contributions from negative-energy states $\Delta g_{\mathrm{int}}(\mathcal{N})$ and positive-energy states $\Delta g_{\mathrm{int}}(\mathcal{P})$.}
\label{Total}
\centering
\begin{ruledtabular}
\begin{tabular}{llllllllllllllll}
& \multicolumn{3}{c}{$2s_{1/2}$} && \multicolumn{3}{c}{$3s_{1/2}$} \\
\cline{2-4} \cline{6-8}
$Z$ & {$\Delta g_{\mathrm{int}}(\mathcal{N})$} & {$\Delta g_{\mathrm{int}}(\mathcal{P})$} & {$\Delta g_{\mathrm{int}}$} && {$\Delta g_{\mathrm{int}}(\mathcal{N})$} & {$\Delta g_{\mathrm{int}}(\mathcal{P})$} & {$\Delta g_{\mathrm{int}}$} \\
\hline
4  & -0.00000249 & 0.00008750(50) & 0.00008501(50) && -0.00000064 & 0.00012132(11) & 0.00012068(11) \\
5  & -0.00000409 & 0.00011208(50) & 0.00010799(50) && -0.00000109 & 0.00017846(12) & 0.00017737(12) \\
6  & -0.00000572 & 0.00013663(47) & 0.00013091(47) && -0.00000155 & 0.00024547(12) & 0.00024392(12) \\
7  & -0.00000735 & 0.00016114(48) & 0.00015379(48) && -0.00000201 & 0.00032230(11) & 0.00032029(11) \\
8  & -0.00000898 & 0.00018574(41) & 0.00017676(41) && -0.00000248 & 0.00040901(10) & 0.00040653(10) \\
9  & -0.00001090 & 0.00021032(38) & 0.00019942(38) && -0.00000295 & 0.00050591(10) & 0.00050297(10) \\
10 & -0.00001222 & 0.00023491(35) & 0.00022269(35) && -0.00000341 & 0.00061255(9)  & 0.00060914(9)  \\
11 & -0.00001383 & 0.00025951(33) & 0.00024569(33) && -0.00000387 & 0.00072916(8)  & 0.00072528(8)  \\
12 & -0.00001542 & 0.00028414(31) & 0.00026872(31) && -0.00000434 & 0.00085584(8)  & 0.00085150(8)  \\
13 & -0.00001702 & 0.00030878(29) & 0.00029177(29) && -0.00000479 & 0.00099274(7)  & 0.00098795(7)  \\
14 & -0.00001860 & 0.00033344(27) & 0.00031485(27) && -0.00000525 & 0.00113945(7)  & 0.00113420(7)  \\
15 & -0.00002015 & 0.00035813(26) & 0.00033797(26) && -0.00000570 & 0.00129628(7)  & 0.00129057(7)  \\
16 & -0.00002178 & 0.00038283(24) & 0.00036105(24) && -0.00000615 & 0.00146336(6)  & 0.00145721(6)  \\
17 & -0.00002326 & 0.00040756(23) & 0.00038431(23) && -0.00000660 & 0.00164053(6)  & 0.00163394(6)  \\
18 & -0.00002478 & 0.00043232(22) & 0.00040754(22) && -0.00000704 & 0.00182807(6)  & 0.00182104(6)  \\
19 & -0.00002629 & 0.00045710(21) & 0.00043081(21) && -0.00000747 & 0.00202584(6)  & 0.00201837(6)  \\
20 & -0.00002779 & 0.00048191(20) & 0.00045412(20) && -0.00000790 & 0.00223372(5)  & 0.00222582(5)  \\
\hline
& \multicolumn{3}{c}{$2p_{1/2}$} && \multicolumn{3}{c}{$2p_{3/2}$} \\
\cline{2-4} \cline{6-8}
$Z$ & {$\Delta g_{\mathrm{int}}(\mathcal{N})$} & {$\Delta g_{\mathrm{int}}(\mathcal{P})$} & {$\Delta g_{\mathrm{int}}$} && {$\Delta g_{\mathrm{int}}(\mathcal{N})$} & {$\Delta g_{\mathrm{int}}(\mathcal{P})$} & {$\Delta g_{\mathrm{int}}$} \\
\hline
4  & -0.00001063 & 0.00011185(13) & 0.00010123(13) &&  0.00000114 & 0.00008096(7)  & 0.000082100(7) \\
5  & -0.00002014 & 0.00015235(15) & 0.00013220(15) &&  0.00000212 & 0.00010550(8)  & 0.000107615(8) \\
6  & -0.00003008 & 0.00019304(14) & 0.00016296(14) &&  0.00000306 & 0.00012982(8)  & 0.000132882(8) \\
7  & -0.00004014 & 0.00023379(13) & 0.00019365(13) &&  0.00000397 & 0.00015406(8)  & 0.000158030(8) \\
8  & -0.00005047 & 0.00027458(12) & 0.00022411(12) &&  0.00000476 & 0.00017826(8)  & 0.000183020(8) \\
9  & -0.00006037 & 0.00031539(11) & 0.00025502(11) &&  0.00000573 & 0.00020244(8)  & 0.000208175(8) \\
10 & -0.00007047 & 0.00035622(10) & 0.00028575(10) &&  0.00000660 & 0.00022663(7)  & 0.000233227(7) \\
11 & -0.00008059 & 0.00039710(10) & 0.00031651(10) &&  0.00000745 & 0.00025082(7)  & 0.000258271(7) \\
12 & -0.00009068 & 0.00043802(9)  & 0.00034733(9)  &&  0.00000829 & 0.00027503(7)  & 0.000283325(7) \\
13 & -0.00010075 & 0.00047898(8)  & 0.00037823(8)  &&  0.00000913 & 0.00029926(7)  & 0.000308396(7) \\
14 & -0.00011081 & 0.00052000(7)  & 0.00040919(7)  &&  0.00000998 & 0.00032351(6)  & 0.000333489(6) \\
15 & -0.00012071 & 0.00056107(7)  & 0.00044036(7)  &&  0.00001089 & 0.00034779(6)  & 0.000358676(6) \\
16 & -0.00013087 & 0.00060221(6)  & 0.00047134(6)  &&  0.00001165 & 0.00037209(6)  & 0.000383746(6) \\
17 & -0.00014086 & 0.00064342(5)  & 0.00050256(5)  &&  0.00001250 & 0.00039643(6)  & 0.000408928(6) \\
18 & -0.00015086 & 0.00068471(5)  & 0.00053385(5)  &&  0.00001333 & 0.00042080(2)  & 0.000434134(2) \\
19 & -0.00016079 & 0.00072607(4)  & 0.00056529(4)  &&  0.00001418 & 0.00044521(5)  & 0.000459393(5) \\
20 & -0.00017072 & 0.00076753(3)  & 0.00059681(3)  &&  0.00001503 & 0.00046966(5)  & 0.000484683(5) \\
\end{tabular}
\end{ruledtabular}
\end{table*}

\section{RESULTS AND DISCUSSION}
\subsection{Interelectronic-interaction corrections}
We present high-precision calculations of the interelectronic-interaction corrections to the Land\'{e} $g$-factors for the $2s_{1/2}$, $2p_{1/2}$, $2p_{3/2}$, and $3s_{1/2}$ states in lithium-like ions ($Z=4 - 20$). Three distinct methods are employed and compared with each other, namely many-body perturbation theory (MBPT), the linearized coupled-cluster singles-doubles (LCCSD) method, and the coupled-cluster singles-doubles (CCSD) method. The interelectronic-interaction correction arising from positive-energy states, $\Delta g_{\mathrm{int}}(\mathcal{P})$, is thereby computed using all three methods to enable robust cross-validation, while the corresponding correction from negative-energy states, $\Delta g_{\mathrm{int}}(\mathcal{N})$, is evaluated within the MBPT framework.

To ensure the reliability of our results, a stringent convergence test with respect to the partial-wave $\ell_{\mathrm{max}}$ expansion is performed for the positive-energy correction $\Delta g_{\mathrm{int}}(\mathcal{P})$, the negative-energy correction $\Delta g_{\mathrm{int}}(\mathcal{N})$, and  the total interelectronic-interaction correction, $\Delta g_{\mathrm{int}} = \Delta g_{\mathrm{int}}(\mathcal{P}) + \Delta g_{\mathrm{int}}(\mathcal{N})$,  within the MBPT, LCCSD, and CCSD frameworks. Taking Ca$^{17+}$ as an example (see Table~\ref{conver}), both $\Delta g_{\mathrm{int}}(\mathcal{N})$ and $\Delta g_{\mathrm{int}}(\mathcal{P})$ are found to converge rapidly with increasing $\ell_{\mathrm{max}}$. For $\ell_{\mathrm{max}} \ge 5$, the values stabilize, and the differences among the results from the three many-body methods fall below $10^{-8}$, confirming excellent convergence. Based on this observation, $\ell_{\mathrm{max}} = 6$ is chosen as a sufficiently high cutoff for all direct calculations.
To quantify the residual uncertainty from the partial-wave truncation, extrapolations to $\ell_{\mathrm{max}} \rightarrow \infty$ are carried out using the sequences $\ell_{\mathrm{max}} = 2$-$3$-$4$-$5$ and $3$-$4$-$5$-$6$ in Eq.~(\ref{extrapolate}). The remarkable agreement between these independently extrapolated values and the direct result at $\ell_{\mathrm{max}}=6$ indicates that the truncation error is negligible at the present level of accuracy. The final recommended values, with the small extrapolation uncertainty given in parentheses, are presented accordingly.

Table~\ref{int_Positive} presents the complete interelectronic-interaction corrections from positive-energy states, $\Delta g_{\mathrm{int}}(\mathcal{P})$, for the $2s_{1/2}$, $2p_{1/2}$, $2p_{3/2}$, and $3s_{1/2}$ states. Across all states, these corrections increase systematically with nuclear charge $Z$, following the trend observed in boron-like systems~\cite{Wang.2025.P109413.v388}. Beyond this typical trend, a pronounced dependence on the specific electronic state is revealed by comparing the results from the MBPT, LCCSD, and CCSD methods. The $2s_{1/2}$ state exhibits the largest variation among the methods, indicating a higher sensitivity to the treatment of correlation, whereas the $2p_{3/2}$ state shows the smallest variation. These results indicate that regardless of whether MBPT, LCCSD, or CCSD (with the correlation level increasing successively) is used, the convergence patterns remain consistent. The CCSD method, which provides the most complete account of singles and doubles excitations within this hierarchy, yields values that represent the converged result for this class of correlation effects. Consequently, we adopt the CCSD results as our final recommended values. A conservative estimate of the associated theoretical uncertainty is taken as the maximum absolute deviation between the CCSD results and those from the MBPT and LCCSD calculations.

The negative-energy state correction, $\Delta g_{\mathrm{int}}(\mathcal{N})$, displays a distinct sensitivity to the electronic state. Its magnitude grows systematically with nuclear charge $Z$, while its sign depends strongly on the specific quantum state. For the $2s_{1/2}$, $3s_{1/2}$, and $2p_{1/2}$ states, $\Delta g_{\mathrm{int}}(\mathcal{N})$ is negative and partially cancels the positive-energy correction $\Delta g_{\mathrm{int}}(\mathcal{P})$, whereas for the $2p_{3/2}$ state, it is positive and enhances the total. This state-dependent sign behavior differs from trends in boron-like systems, where all $2p_j$ states yield a negative sign~\cite{Wang.2025.P109413.v388}. Meanwhile, $\Delta g_{\mathrm{int}}(\mathcal{N})$ is typically one or two orders of magnitude smaller than $\Delta g_{\mathrm{int}}(\mathcal{P})$, reaching three orders of magnitude smaller for the $3s_{1/2}$ state. As a result, $\Delta g_{\mathrm{int}}(\mathcal{P})$ affects the third significant digit of the $g$-factor for $3s_{1/2}$, whereas $\Delta g_{\mathrm{int}}(\mathcal{N})$ influences only the sixth digit, making its inclusion essential for achieving seven significant digits of accuracy for such states. However, this pattern breaks down for the $2p_{1/2}$ state, which shows a markedly enhanced sensitivity to the negative-energy contribution. Here, the relative importance of $\Delta g_{\mathrm{int}}(\mathcal{N})$ is vastly enhanced. For high-$Z$ ions, its magnitude becomes comparable to that of $\Delta g_{\mathrm{int}}(\mathcal{P})$, and its effect emerges as early as the fourth significant digit. This underscores the critical and highly state-specific role of the negative-energy correction in precision $g$-factor theory. The total interelectronic-interaction correction $\Delta g_{\mathrm{int}}$ is provided in Table~\ref{Total}, with its uncertainty dominated by the treatment of electron correlation in $\Delta g_{\mathrm{int}}(\mathcal{P})$.

\begin{figure}
            \centering
            \includegraphics[width=0.99\columnwidth]{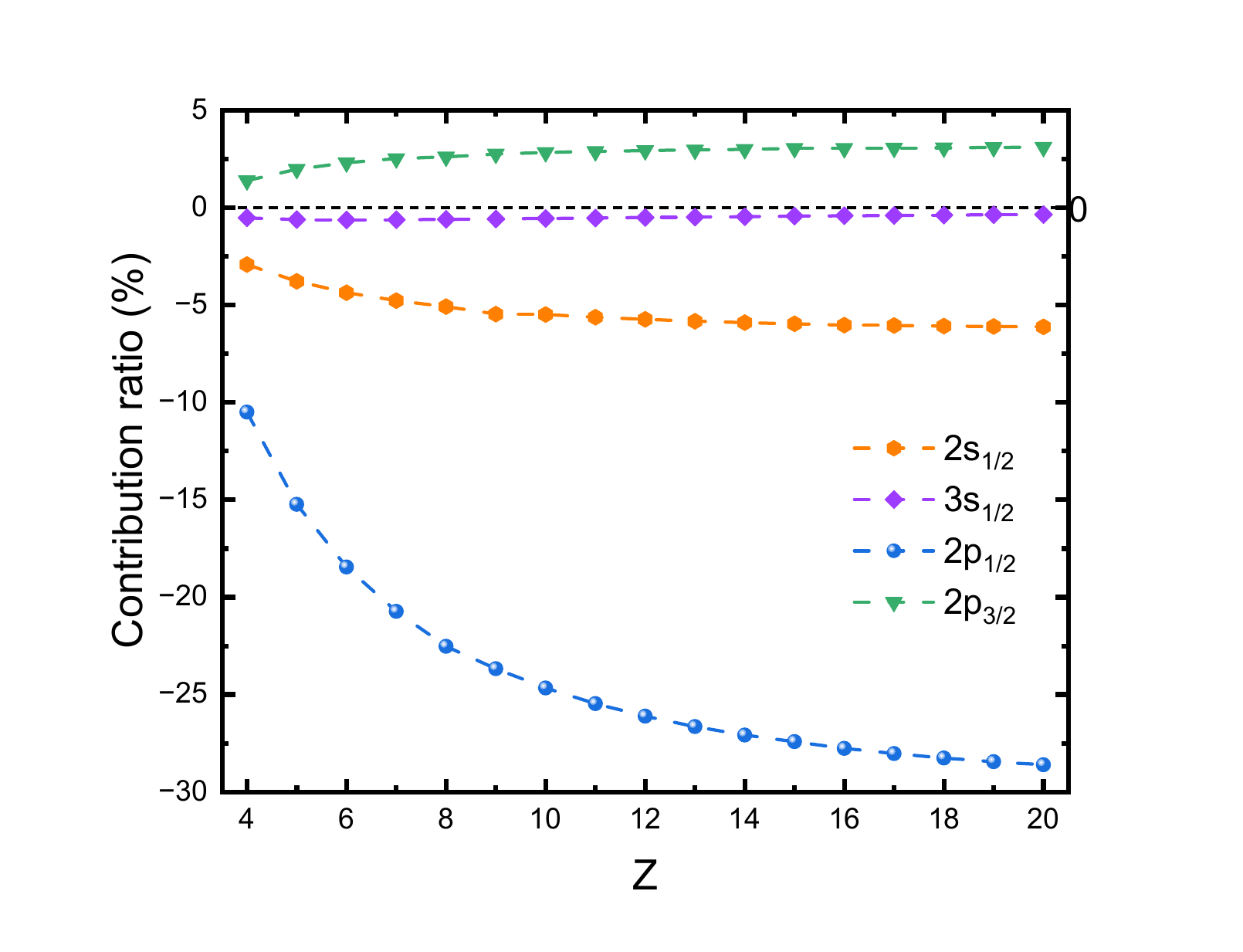}
            \caption{The contribution from the negative-energy states to the interelectronic-interaction corrections.}
            \label{fig:negative}
\end{figure}

Figure~\ref{fig:negative} quantifies the relative contribution of $\Delta g_{\mathrm{int}}(\mathcal{N})$, demonstrating its essential role in precision theory. For the $2p_{1/2}$ state, this relative contribution increases most strongly with $Z$, reaching about 30\% for $Z = 20$. Therefore, leaving it out for this state would lead to a large error. Notable contributions are also seen for the $2s_{1/2}$ state (up to 6\%) and the $2p_{3/2}$ state (up to 3\%). Even for the $3s_{1/2}$ state, where the relative effect is 0.6\%, the contribution cannot be ignored in high-precision studies. These results clearly show that including negative-energy state contributions is important for accurate $g$-factor calculations.

\begin{sidewaystable}
\normalfont
\caption{\label{int_3s0.5}The ground state Land\'{e} $g$-factor of lithium-like ion with $Z = 4 - 20$. $\Delta g_{\mathrm{int}}$ denotes the interelectronic-interaction correction, where $\Delta g_{\mathrm{int}}^{\mathrm{present}}$ is the current value. $\Delta g_{\mathrm{QED}}$, $\Delta g_{\mathrm{rec}}$, and $\Delta g_{\mathrm{NS}}$ represent the QED, nuclear recoil, and nuclear size corrections, respectively. $ g_{\mathrm{th}}$ and $g_{\mathrm{expt}}$ are the theoretical and experimental $g$-factors. The values in parentheses indicate the uncertainties.
}
\begin{ruledtabular}
\tiny
\setlength{\tabcolsep}{1pt}
\begin{tabular}{lllllllllllll}
                           effect        &$Z = 6$            &$Z = 8$            &$Z = 10$            & $Z = 12$     &  $Z = 14$       & $Z = 16$      &  $Z = 18$   & $Z = 20$         \\
 \hline
$g_D$                         & 1.99968030                                          & 1.99943138                                           & 1.99911100                                      & 1.99871889                                     &   1.99825475                                                 &  1.99771819                                     & 1.99710878                                      &   1.99642601                                          \\
$\Delta g_{\rm{int}}^{\rm{present}}$   & 0.00013091(47)                               & 0.00017676(41)                                       & 0.00022269(35)                                  & 0.00026872(31)                                 &    0.00031485(27)                                            &  0.00036105(24)                                 & 0.00040754(22)                                  &   0.00045412(20)                                      \\
$\Delta g_{\rm{int}}$         & 0.000130758(19)\cite{Glazov.2004.V70.p062104}       &  0.000176658(30)\cite{Glazov.2004.V70.p062104}       &  0.000222628(42)\cite{Glazov.2004.V70.p062104}  & 0.000268703(55)\cite{Glazov.2004.V70.p062104}  &    0.0003148098(22)\cite{Kosheleva.2022.V128.p103001}        &   0.00036124(9)\cite{Glazov.2004.V70.p062104}    &  0.00040775(12)\cite{Glazov.2004.V70.p062104}   &   0.0004542910(24)\cite{Kosheleva.2022.V128.p103001}  \\
                            & 0.000130758(19)\cite{Moskovkin.2008.V77.p063421}    &  0.00017666(3)\cite{Moskovkin.2008.V77.p063421}      &                                                 &                                                &    0.0003148118(27)\cite{Glazov.2019.V123.p173001}           &   0.00036124(9)\cite{Moskovkin.2008.V77.p063421} &                                                 &   0.00045445(14)\cite{Moskovkin.2008.V77.p063421}     \\
                            &                                                     &                                                      &                                                 &                                               &    0.000314809(6)\cite{Volotka.2014.V112.p253004}             &                                                 &                                                 &   0.000454290(9)\cite{Volotka.2014.V112.p253004}      \\
                            &                                                     &                                                      &                                                 &                                                &                                                              &                                                 &                                                 &   0.00045445(14)\cite{Glazov.2004.V70.p062104}       \\
$\Delta g_{\rm{QED}}$           & 0.002319417(6)\cite{Glazov.2004.V70.p062104}        & 0.002319550(12)\cite{Glazov.2004.V70.p062104}        & 0.002319739(21)\cite{Glazov.2004.V70.p062104}   & 0.002319992(32)\cite{Glazov.2004.V70.p062104}  &    0.0023202857(17)\cite{Kosheleva.2022.V128.p103001}        &  0.002320704(7)\cite{Glazov.2004.V70.p062104}   &  0.002321166(9)\cite{Glazov.2004.V70.p062104}   &   0.0023216601(17)\cite{Kosheleva.2022.V128.p103001}  \\
                            & 0.002319417(6)\cite{Moskovkin.2008.V77.p063421}     & 0.002319549(12)\cite{Moskovkin.2008.V77.p063421}     &                                                 &                                                 &                                                             &  0.00232070(6)\cite{Moskovkin.2008.V77.p063421} &                                                 &                                                       \\
$\Delta g_{\text{rec}}$          & 0.00000001\cite{Glazov.2004.V70.p062104}            & 0.000000017\cite{Glazov.2004.V70.p062104}            & 0.000000025\cite{Glazov.2004.V70.p062104}       & 0.000000032\cite{Glazov.2004.V70.p062104}      &    0.0000000436\cite{Kosheleva.2022.V128.p103001}            & 0.000000046(1)\cite{Glazov.2004.V70.p062104}    & 0.000000048(1)\cite{Glazov.2004.V70.p062104}    &   0.0000000662\cite{Kosheleva.2022.V128.p103001}      \\
                            & 0.000000009\cite{Moskovkin.2008.V77.p063421}        & 0.000000016\cite{Moskovkin.2008.V77.p063421}         &                                                 &                                                &                                                              &  0.000000045(1)\cite{Moskovkin.2008.V77.p063421}&                                                &                                                       \\
$\Delta g_{\rm{NS}}$            & 0                                                   & 0                                                    & 0.000000001\cite{Glazov.2004.V70.p062104}       & 0.000000001\cite{Glazov.2004.V70.p062104}      &     0.000000005\cite{Glazov.2004.V70.p062104}                &  0.000000009\cite{Glazov.2004.V70.p062104}      &                                                 &   0.000000014\cite{Volotka.2014.V112.p253004}         \\
                            & 0                                                   & 0                                                    &                                                 &                                                &     0.0000000026\cite{Glazov.2019.V123.p173001}              &  0.000000005\cite{Moskovkin.2008.V77.p063421}   &                                                 &                                                         \\
$g_{\rm {th}}^{\rm{present}}$   & 2.00213064(47)                                      & 2.00192771(43)                                       & 2.00165345(41)                                  & 2.00130764(45)                                 &     2.00088993(32)                                           &  2.00040041(25)                                 & 1.99983798(24)                                  &   1.99920187 (23)                                     \\
$g_{\rm {th}}$                  & 2.002130485(19)\cite{Glazov.2004.V70.p062104}       & 2.001927604(32)\cite{Glazov.2004.V70.p062104}        & 2.001653389(47)\cite{Glazov.2004.V70.p062104}   &  2.001307619(64)\cite{Glazov.2004.V70.p062104} &     2.0008898924(28)\cite{Kosheleva.2022.V128.p103001}       &  2.00040019(11)\cite{Glazov.2004.V70.p062104}   & 1.99983775(14)\cite{Glazov.2004.V70.p062104}    &   1.99920224(17)\cite{Glazov.2004.V70.p062104}       \\
                            & 2.0021304588(10)\cite{Yerokhin.2021.V104.p022814}   & 2.0019275585(10)\cite{Yerokhin.2021.V104.p022814}    &                                                 &                                                &    2.0008898937(17)\cite{Yerokhin.2021.V104.p022814}         &  2.00040018(11)\cite{Moskovkin.2008.V77.p063421}&                                                 &   1.9992020426(29)\cite{Kosheleva.2022.V128.p103001}   \\
                            & 2.0021304573615(13)\cite{Yerokhin.2017.V95.p062511} & 2.0019275584615(14)\cite{Yerokhin.2017.V95.p062511}  &                                                 &                                                &    2.0008898944(34)\cite{Glazov.2019.V123.p173001}           &                                                 &                                                 &   1.9992020529(27)\cite{Yerokhin.2021.V104.p022814}    \\
                            &                                                     &                                                      &                                                 &                                                &    2.000889892(8)\cite{Volotka.2014.V112.p253004}            &                                                 &                                                 &   1.999202041(13)\cite{Volotka.2014.V112.p253004}       \\
 $g_{\rm {expt}}$ \\           &                                                     &                                                      &                                                 &                                                &     2.00088988845(14)\cite{Glazov.2019.V123.p173001}         &                                                 &                                                 &    1.9992020405(11)\cite{Florian.2016.V7.p10246}      \\
                            &                                                     &                                                      &                                                 &                                                &     2.0008898884(19)\cite{Wagner.2013.V110.p033003}          &                                                 &                                                 &                                                      \\
\hline     \\
                            &   $Z = 4$                                           &$Z=5$                                               &  $Z=7$                                              &    $Z=9$                                       &   $Z=11$                                                     &   $Z=13$                                       &   $Z=15$                                         &   $Z=17 $                       &   $Z=19$           \\
$g_{\rm{D}}$                    &  1.99985796                                         &1.99977803                                          &  1.99956476                                         &  1.99928014                                    &  1.99892393                                                  &   1.99849585                                   &   1.99799555                                     &   1.99742262                  &   1.99677660        \\
$\Delta g_{\rm{int}}$           &  0.00008501(50)                                     &0.00010799(50) 	                                   &  0.00015379(48)                                     &  0.00019942(38)                                &	0.00024569(33)                                               &  0.00029177(29)                                &   0.00033797(26)                                 &   0.00038431(23)               &   0.00043081(21)   \\
$\Delta g_{\rm{QED}}$           &  0.002320015(6)                                     &0.002320047(6)                                      &  0.002320150(12)                                    &  0.002320308(21)                               & 0.002320526(32)                                              &  0.002320810(17)                               &   0.002321163(7)                                 &   0.002321591(9)              &   0.002322099(11)   \\
$\Delta g_{\text{rec}}$         &  $-$0.000000020                                       &0.000000000                                         &  0.000000015                                        &  0.000000020                                   & 0.000000028                                                  &   0.000000038                                  &   0.000000045(1)                                 &   0.000000047(1)              &   0.000000053       \\
$g_{\rm {th}}$                  &  2.00226297(50)                                      &2.00220606(50)                                      &  2.00203871(49)                                     &  2.00179988(43)                                &  2.00149017(46)                                              &   2.00110846(34)                               &   2.00065473(27)                                 &   2.00012857(25)              &   1.99952956(24)    \\

\end{tabular}
\end{ruledtabular}
\end{sidewaystable}

\begin{table*}
\caption{\label{int_2pj}The Land\'{e} $g$-factor and theoretical contributions for the first excited state of lithium-like ion with $Z = 4 - 20$. The values in parentheses indicate the uncertainties.
}
\begin{ruledtabular}
\scriptsize
\setlength{\tabcolsep}{1pt}
\begin{tabular}{llllllllll}
\multirow{2}{*}{effect} & \multicolumn{6}{c}{ $2p_{1/2}$ }  \\
\cline{2-7}
                                   &$Z=10$         &  $Z=12$         & $Z=14$          & $Z=16$          &  $Z=18$          & $Z=20$      \\
 \hline
$g_D$                                         &  0.66577766 	      &  0.66538556 	      & 0.66492142 	      & 0.66438486 	     &  0.66377545      &	0.66309268         \\
$\Delta g_{\rm{int}}^{\rm{present}}$              &  0.00028575(10)     &  0.00034733(9)      & 0.00040919(7)     & 0.00047134(6)    &  0.00053385(5)   & 0.00059681(3)      \\
$\Delta g_{\rm{int}}$\cite{zinenko.2025.P.v}    &  0.0002855(19)      &  0.0003469(16)      & 0.0004085(14)     & 0.0004704(13)    &  0.0005325(12)   & 0.0005949(11)      \\
$\Delta g_{\rm{QED}}$\cite{zinenko.2025.P.v}    &$-$0.00077181(6)     &$-$0.00077114(9)     &$-$0.00077029(11)  &$-$0.00076925(13) &$-$0.000768(15)   & $-$0.00076654(18)  \\
$\Delta g_{\text{rec}}$\cite{zinenko.2025.P.v}   &$-$0.0000244(21)     &$-$0.0000199(14)     &$-$0.0000168(11)   &$-$0.00001451(84) &$-$0.00001150(60) & $-$0.00001140(54)  \\
$g_{\rm {th}}^{\rm {present}}$                  &  0.66526720(24)     &  0.66494185(19)     & 0.66454352(17)    &  0.66407244(85)  &  0.66352979(62)  & 0.66291155(57)     \\
$g_{\rm{th}}$\cite{zinenko.2025.P.v}            &  0.6652670(28)      &  0.6649414(21)      & 0.6645428(18)     &  0.6640715(16)   &  0.6635285(14)   & 0.6629096(12)     \\
$g_{\rm{th}}$\cite{Yan.2002.V66.p}            &0.66526595102(5)    &0.66494105221(5)    &    \\
\hline
	                                            &   $Z = 4$           &  $Z = 5$            &   $Z = 6$           &     $Z = 7$       &  $Z = 8$         &  $Z = 9$          \\
$g_D$	                                        & 0.66652463          &  0.66644469         &   0.66634697        &  0.66623143       &  0.66609805      &  0.66594680       \\
$\Delta g_{\rm{int}}$                           & 0.00010123(13)      &  0.00013220(15)     &   0.00016296(14)    &  0.00019365(13)   &  0.00022411(12)  &  0.00025502(11)   \\
$\Delta g_{\rm{QED}}$	                          & $-$0.00077258(6)    &  $-$0.000772575(6)   &  $-$0.00077252(6)   & $-$0.000772415(6) &  $-$0.000772261(6)& $-$0.000772057(6)\\
$\Delta g_{\text{rec}}$	                        & 0.00008910(21)      &  0.00003405(21)     &   0.00000133(21)    & $-$0.00001608(21) &  $-$0.00002361(21)& $-$0.00002534(21) \\
$g_{\rm {th}}^{\rm {present}}$	                & 0.66594237(25)          &  0.66583837(26)     &   0.66573874(26)    &  0.66563659(25)   &  0.66552629(25)   &  0.66540442(24)  \\
$g_{\rm{th}}$\cite{Yan.2002.V66.p}              & 0.6657835556(3)   & 0.66575107987(7)    &  0.66569085558(3)   &0.66561402014(3)   &0.66551703820(4)   & 0.66540233507(7) \\
	                                            &   $Z = 11$	        &    $Z = 13$         &  $Z = 15$         &   $Z = 17$       &	  $Z = 19$      &                   \\
$g_D$	                                        &  0.66559059 	      & 0.66516252 	        & 0.66466222 	      &  0.66408929 	   &  0.66344327      &                   \\
$\Delta g_{\rm{int}}$	                          &  0.00031651(10)     & 0.00037823(8)       & 0.00044036(7)     &  0.00050256(5)   &  0.00056529(4)   &                   \\
$\Delta g_{\rm{QED}}$	                          &$-$0.00077150(9)     &$-$0.00077075(11)    &$-$0.00076979(13)  &$-$0.00076864(15) &$-$0.00076729(18) &                   \\
$\Delta g_{\text{rec}}$	                        &$-$0.00002192(14)    &$-$0.00001794(11)    &$-$0.00001555(84)  &$-$0.00001294(60) &$-$0.00001041(54) &                   \\
$g_{\rm {th}}^{\rm {present}}$	                &  0.66511369(19)     & 0.66475206(17) 	    & 0.66431723(85) 	  &  0.66381026(62)  &  0.66323085(57)  &                   \\
$g_{\rm{th}}$\cite{Yan.2002.V66.p}            & 0.66511358663(6)&\\
\hline
& & & $2p_{3/2}$\\
 \hline
                                              &  $Z=10$             &  $Z=12$             & $Z=14$             & $Z=16$           &  $Z=18$            & $Z=20$             \\
$g_{\rm{D}}$                                    &  1.3326231  	      &  1.3323104   	      &    1.3319408  	   &  1.3315141 	    &  1.3310304  	     &  1.3304895       \\
$\Delta g_{\rm{int}}^{\rm{present}}$              &  0.00023323(7)      &  0.00028332(7)      &    0.00033349(6)   &  0.00038375(6)   &  0.00043413(2)     &  0.00048468(5)    \\
$\Delta g_{\rm{int}}$\cite{zinenko.2025.P.v}    &  0.0002332(15)      &  0.0002831(12)      &    0.0003330(11)   &  0.0003829(10)   &  0.0004328(9)      &  0.0004828(8)    \\
$\Delta g_{\rm{QED}}$\cite{zinenko.2025.P.v}    &  0.00077682(8)      &  0.00077755(10)     &    0.00077844(12)  &  0.00077951(14)  &  0.00078077(17)    &  0.00078227(19)    \\
$\Delta g_{\text{rec}}$\cite{zinenko.2025.P.v}   &$-$0.0000122(11)     &$-$0.00000998(75)    &  $-$0.00000842(55) & $-$0.00000729(42)&  $-$0.00000578(30) & $-$0.00000574(27)  \\
$\Delta g_{\rm{int}}^{\rm{present}}$              &  1.33362093(15)     &  1.33336131(76)     &    1.33304430(57)  &	1.33267010(45)  &  1.33223951(35)    &  1.33175068(33)   \\
$g_{\rm{th}}$\cite{zinenko.2025.P.v}             &   1.3336209(19)     &  1.3333611(14)      &    1.3330438(12)   &  1.3326693(11)   &  1.3322382(10)     &  1.3317488(9)    \\
$g_{\rm{th}}$\cite{Yan.2002.V66.p}            & 1.33361701465(4)    & 1.33335676135(4)  \\
\hline
                                              &    $Z=4$            & $Z = 5$             &   $Z = 6$           &   $Z = 7$          &  $Z = 8$         &  $Z = 9$          \\
$g_{\rm{D}}$	                                  & 1.3332197  	        & 1.3331558 	        &  1.3330777 	        &  1.3329854 	       &  1.3328788 	    &  1.3327581         \\
$\Delta g_{\rm{int}}$	                          & 0.0000821(8)        & 0.0001076(8)        &  0.0001329(8)       &  0.0001580(8)      &  0.0001830(8)    &  0.0002082(8)         \\
$\Delta g_{\rm{QED}}$	                          & 0.0007759(8)        & 0.0007759(8) 	      &  0.0007760(8) 	    &  0.0007761(8) 	   &  0.0007763(8) 	  &  0.0007765(8)         \\
$\Delta g_{\text{rec}}$	                        & 0.0000511(11)       & 0.0000209(11) 	    &  0.0000028(11) 	    & $-$0.0000070(11) 	 & $-$0.0000114(11) & 	$-$0.0000125(11)    \\
$g_{\rm {th}}^{\rm {present}}$                  & 1.3341288(16)       & 1.3340602(16) 	    &  1.3339893(16) 	    &  1.3339125(16) 	   &  1.3338268(16) 	&  1.3337302(16)        \\
$g_{\rm{th}}$\cite{Yan.2002.V66.p}            & 1.3340406063(1)                     &  1.33401044835(5)   & 1.33396092650(3)    & 1.33389768859(3)   &1.33381900332(4)   & 1.33372610185(6)  \\
                                              & $Z=11$              &	$Z=13$ 	            & $Z=15$             & $Z =17$          &  $Z = 19$          &                  \\
$g_{\rm{D}}$	                                  & 1.3324739 	        & 1.3321327 	        &   1.3317346 	     & 1.3312794 	      & 1.3307671          &                  \\
$\Delta g_{\rm{int}}$	                          & 0.0002583(7) 	      & 0.0003084(7) 	      &   0.0003587(6) 	   & 0.0004089(6) 	  & 0.0004594(5)       &                  \\
$\Delta g_{\rm{QED}}$	                          & 0.0007772(10) 	    & 0.0007780(12) 	    &   0.0007789(14) 	 & 0.0007801(17) 	  & 0.0007815(19)      &                  \\
$\Delta g_{\text{rec}}$	                        &$-$0.0000110(75)     &$-$0.0000090(55)     &  $-$0.0000078(42)  &$-$0.0000065(30)  &$-$0.0000052(27)    &                  \\
$g_{\rm {th}}^{\rm {present}}$                  &	 1.3334983(76) 	    & 1.3332101(57) 	    &   1.3328644(45) 	 & 1.3324620(35) 	  & 1.3320028(33)      &                  \\
$g_{\rm{th}}$\cite{Yan.2002.V66.p}            &1.33349459813(5)  &    \\
\end{tabular}
\end{ruledtabular}
\end{table*}

\subsection{Land\'{e} $g$-factor of ground state}
Table~\ref{int_3s0.5} presents the total $g$-factor for the ground state $2s_{1/2}$ and its individual contributions. The total $g$-factor is obtained by combining our calculated interelectronic-interaction correction, $\Delta g_{\mathrm{int}}^{\mathrm{present}}$, with corrections for QED ($\Delta g_{\mathrm{QED}}$), nuclear recoil ($\Delta g_{\mathrm{rec}}$), and nuclear size ($\Delta g_{\mathrm{NS}}$), whose values are adopted from the literature~\cite{Glazov.2004.V70.p062104,Moskovkin.2008.V77.p063421, Kosheleva.2022.V128.p103001,Glazov.2019.V123.p173001}. Their precise evaluation requires rigorous QED methods beyond the scope of the present many-body approach. For most ions, the nuclear corrections are taken from the comprehensive dataset of Glazov $et~al$.~\cite{Glazov.2004.V70.p062104}, which covers a broad range of $Z$. For the well-studied ions Si$^{11+}$ and Ca$^{17+}$, we employ the more recent and precise results of Kosheleva $et~al$.~\cite{Kosheleva.2022.V128.p103001}, which include higher-order effects such as screened QED and two-photon exchange. The table also provides a direct comparison of our $\Delta g_{\mathrm{int}}^{\mathrm{present}}$ with previous theoretical values from the literature~\cite{Glazov.2004.V70.p062104, Kosheleva.2022.V128.p103001,Moskovkin.2008.V77.p063421, Volotka.2014.V112.p253004,Glazov.2019.V123.p173001}.

A direct comparison of the interelectronic-interaction corrections, presented in Table~\ref{int_3s0.5}, validates our computational methodology. Our recommended $\Delta g_{\mathrm{int}}^{\mathrm{present}}$ values show excellent agreement with those of Glazov $et~al$.~\cite{Glazov.2004.V70.p062104}, matching at least three significant digits across the $Z$-range studied. For most ions, this agreement ensures that the total $g$-factor achieves an accuracy of at least seven significant digits, which is consistent with established theoretical and experimental benchmarks~\cite{Yerokhin.2017.V95.p062511,Yerokhin.2021.V104.p022814,Moskovkin.2008.V77.p063421, Volotka.2014.V112.p253004,Kosheleva.2022.V128.p103001,Glazov.2019.V123.p173001, Wagner.2013.V110.p033003,Florian.2016.V7.p10246}. An evaluation of the complete set of corrections clarifies their respective roles. The QED correction $\Delta g_{\mathrm{QED}}$ provides the dominant contribution, followed by our interelectronic-interaction correction $\Delta g_{\mathrm{int}}$, the latter of which typically affects the fourth decimal place of the total. A slight deviation is noted for calcium ions ($Z = 20$). While the agreement for $\Delta g_{\mathrm{int}}$ remains at the level of three significant digits, the final $g$-factor accuracy is maintained at six significant digits, in line with the benchmark. This reflects the increased sensitivity of higher-$Z$ systems to nuclear effects, where the associated uncertainties can become the limiting factor for overall precision.
Overall, when all uncertainties are considered, our results are fully consistent with previous studies, confirming the reliability of our theoretical framework across the entire $Z$-range.

In addition to validating existing results, this work provides missing data for odd-$Z$ nuclei. As shown in Table~\ref{int_3s0.5}, previous theoretical studies on lithium-like ions mainly focused on even-$Z$ atomic nuclei, resulting in a scarcity of experimental and theoretical reference data for odd-$Z$ lithium-like ions. Therefore, based on the known high-precision data of even-$Z$ ions~\cite{Glazov.2004.V70.p062104,zinenko.2025.P.v}, we use a smooth interpolation method to obtain the QED correction ($\Delta g_{\mathrm{QED}}$) and the nuclear recoil correction ($\Delta g_{\mathrm{rec}}$). In the isoelectronic sequence, the $\Delta g_{\mathrm{QED}}$ depends primarily on the atomic number $Z$ and varies smoothly and monotonically with $Z$; thus, the interpolation results based on even-$Z$ ions can be reliably extended to odd-$Z$ ions~\cite{Wang.2025.P109413.v388}. In contrast, the $\Delta g_{\mathrm{rec}}$ depends on both $Z$ and the nuclear mass number $A$, and the discrete nature of $A$ may slightly reduce the interpolation accuracy~\cite{Wang.2025.P109413.v388}. Nevertheless, within the present $Z$ range, the variation of $A$ remains sufficiently regular, and the interpolated values still meet the required accuracy of this work. The uncertainty of each interpolated value is conservatively taken as that of the adjacent even-$Z$ ion. For the excited states, where data are also missing for $Z < 10$, the uncertainty is conservatively taken as the maximum uncertainty among the known ions.

\subsection{Land\'{e} $g$-factor of excited states}
While high-precision benchmarks are well established for the ground state, corresponding data for excited states are less precise. This stems from the considerable challenge in extending the precise methods applicable to few-body systems to excited states. In this work, we report new calculations for the $2p_{1/2}$ and $2p_{3/2}$ excited states across the $Z=4$-$20$ range. The total $g$-factors and their individual contributions are presented in Table~\ref{int_2pj}. For these states, the QED ($\Delta g_\text{QED}$) and nuclear recoil ($\Delta g_\text{rec}$) corrections are taken from the recent work by Zinenko $et~al$., which employs a rigorous few-body approach~\cite{zinenko.2025.P.v}. Our calculated interelectronic-interaction correction, $\Delta g_{\text{int}}^{\text{present}}$, is compared with the results from this reference and with the earlier values of Yan $et~al$.~\cite{Yan.2002.V66.p}.

Our calculated $\Delta g_{\text{int}}^{\text{present}}$ matches the results of Zinenko $et~al$~\cite{zinenko.2025.P.v} and is consistently more precise, typically by about one significant digit. For the $2p_{1/2}$ state, where the interelectronic-interaction dominates, the two sets of values agree to within $2$-$3$ significant digits across the $Z$-series, with closer agreement at lower $Z$. For the $2p_{3/2}$ state, the agreement ranges from $2$ to $4$ significant digits, also improving at lower $Z$. In contrast to the $2p_{1/2}$ case, the QED contribution here is larger, which influences the comparison. All comparisons are consistent within the quoted uncertainties. Overall, the agreement for both states is excellent and systematically improves with decreasing $Z$, confirming the consistency and reliability of our correlation treatment.

The excellent agreement for $\Delta g_{\text{int}}$ leads to consistent total $g$-factors. Compared directly to the results of Zinenko $et~al$.~\cite{zinenko.2025.P.v} and Yan $et~al$.~\cite{Yan.2002.V66.p}, our values demonstrate high precision for both excited states. For the $2p_{1/2}$ state, our total $g$-factor agrees with that of Zinenko $et~al$. to 4-6 significant digits and with Yan $et~al$. to 3-6 digits. The agreement is even closer for the $2p_{3/2}$ state, reaching 5 to 8 significant digits with Zinenko $et~al$. and 4-6 digits with Yan $et~al$.
These results demonstrate that for excited states, the present many-body method achieves accuracy comparable to, and often slightly exceeding, that of specialized high-precision few-body methods where direct benchmarks are available.

The observed systematic differences between the referenced studies can arise from their distinct treatment to electron correlation and nuclear corrections. The work of Yan $et~al.$~\cite{Yan.2002.V66.p} provides a valuable description of correlation but omits certain higher-order nuclear corrections (e.g., relativistic recoil terms of order $\alpha^2 m/M$ and $\alpha^3$ radiative contributions), which limits its final accuracy, especially in high-$Z$, QED-dominated regimes. On the other hand, the study by Zinenko $et~al.$~\cite{zinenko.2025.P.v} incorporates a more systematic set of nuclear corrections, but their treatment to electron correlation is truncated at second order, restricting the precision of their $\Delta g_{\text{int}}$, particularly for the correlation-sensitive $2p_{1/2}$ state. In the present work, we adopt a complementary strategy by employing the CCSD method to treat electron correlation more systematically while synthesizing the most accurate nuclear correction data. Consequently, our results demonstrate good consistency for different states across the entire Z-range studied, remaining in agreement with existing studies within their reported uncertainties. Thus, the present work establishes a unified treatment that combines accurate electron correlation with state-of-the-art nuclear corrections, enabling high-precision predictions for excited states across the studied $Z$-range, with a methodology that is promising for extension to a broader spectrum of systems.

\section{Summary}
In this work, we have systematically calculated the Land\'{e} $g$-factors for the ground state $2s_{1/2}$ and the low-lying excited states $2p_{1/2}$, $2p_{3/2}$, and $3s_{1/2}$ in lithium-like ions across the nuclear charge range $Z = 4$ to $20$ using the MBPT, LCCSD, and CCSD methods. Our approach combines the all-order coupled-cluster method for evaluating contributions from positive-energy states with third-order perturbation theory for assessing negative-energy state contributions. We find that the contributions from negative-energy states exhibit a remarkable state-dependence in both magnitude and sign. For the $2p_{1/2}$ state, this contribution reaches approximately 30\% of the interelectronic-interaction correction and becomes comparable in magnitude to that from positive-energy states at high $Z$, while maintaining a negative sign that leads to partial cancellation. In contrast, the $2p_{3/2}$ state shows a positive contribution that is one order of magnitude smaller than that of the positive-energy state. The $2s_{1/2}$ state exhibits a contribution of up to 6\%, significant at the fifth decimal place, and even the smallest contribution, found in the $3s_{1/2}$ state, appears at the sixth decimal place and is non-negligible.
These results clearly confirm that the negative-energy state correlations play an indispensable role in the high-precision determination of the $g$-factor, regardless of whether it is for the $s$-state or $p$-state.

Based on the above results, we present the total $g$-factors by integrating precise QED and nuclear corrections~\cite{Glazov.2004.V70.p062104,zinenko.2025.P.v}. For the ground state, our results show excellent agreement with established benchmarks, thereby validating our computational methodology. For the excited $2p_j$ states, our treatment incorporates a more comprehensive description of electron correlations, potentially yielding more accurate values than prior studies. However, for high-$Z$ systems, frequency-dependent Breit effects become significant and warrant future investigation.

The present results provide benchmark results for many-body calculations of Land\'{e} $g$-factors in heavy atomic systems and can serve as a reference for future theoretical studies using relativistic many-body methods. We quantitatively reveal the significant state dependence of the contributions from negative-energy states and demonstrate that these contributions are indispensable for achieving spectroscopic accuracy. Furthermore, the computational framework developed in this work achieves an accuracy comparable to that of specialized few-body methods while remaining rigorously applicable to many-electron systems, thereby providing an effective tool for high-precision, scalable calculations of more complex ions in the future.

\begin{acknowledgments}
We are grateful to Yong-Jun
Cheng for reading our manuscript. This work was supported by the National Natural Science Foundation of China (Grants No. 12504295, No. 12174268).
\end{acknowledgments}

\bibliography{Li}

\end{document}